\documentclass[
12pt,
amsmath,
superscriptaddress,
floatfix,
noshowkeyws,
notitlepage,
noshowpacs,
]{revtex4-1}

\usepackage{graphicx}
\usepackage{amsmath}
\usepackage{comment}
\usepackage{amssymb}

\usepackage{subfigure}
\usepackage{tikz}

\usepackage{subfigure}
\usepackage{color}
\usepackage{soul}

\begin{document}

\title{Controlling turbulent drag across electrolytes using electric fields} 

\author{Rodolfo Ostilla-M\'{o}nico}
\email{rostillamonico@g.harvard.edu}
\affiliation{Harvard John A. Paulson School of Engineering and Applied Sciences, Harvard University, Cambridge, MA 02138}

\author{Alpha A. Lee}
\email{alphalee@g.harvard.edu}
\affiliation{Harvard John A. Paulson School of Engineering and Applied Sciences, Harvard University, Cambridge, MA 02138}

\begin{abstract}
Reversible \emph{in operando} control of friction is an unsolved challenge crucial to industrial tribology. Recent studies show that at low sliding velocities, this control can be achieved by applying an electric field across electrolyte lubricants. However, the phenomenology at high sliding velocities is yet unknown. In this paper, we investigate the hydrodynamic friction across electrolytes under shear beyond the transition to turbulence. We develop a novel, highly parallelised, numerical method for solving the coupled Navier-Stokes Poisson-Nernest-Planck equation. Our results show that turbulent drag cannot be controlled across dilute electrolyte using static electric fields alone. The limitations of the Poisson-Nernst-Planck formalism hints at ways in which turbulent drag could be controlled using electric fields. 
\end{abstract}

\makeatother
\maketitle

\section{Introduction} 

It is estimated that one-fifth of all energy produced globally is lost to friction \cite{holmberg2011global}. In industrialised countries such as the UK and US, advances in tribology could save up to 1.4\% of gross national product \cite{jost2005commentary}. Although systematic studies of friction date back at least to the time of Leonardo da Vinci \cite{hutchings2016leonardo}, a molecular understanding of friction remains a challenge because friction is a highly non-linear and far-from-equilibrium phenomenon that is intimately dependent on nanoscale contacts between surfaces \cite{urbakh2004nonlinear,urbakh2010nanotribology,vanossi2013colloquium}.

Empirically, it is known since antiquity that sandwiching a liquid between two surfaces could reduce friction between the surfaces. The well-known Stribeck Curve shows that lubrication is a distinctly multiscale problem --- at slow sliding speeds/narrow surface separations it is the interactions between surface asperities that dominate friction (boundary friction), whereas at high sliding velocity elastohydrodynamic effects of the lubricant becomes important \cite{persson2013sliding}.  The challenge lies in finding the optimal lubricant for a given pair of surfaces and condition. 

The classic lubricant is ``oil'', i.e. long chain hydrocarbons. In recent years, ionic liquids, molten salts at room temperature, are being used as lubricants for mechanical parts such as engines and ball bearings. There are many physiochemical properties of ionic liquids that make them desirable lubricants. For example, most ionic liquids have negligible volatility, high thermal stability and are non-flammable \cite{ye2001room,zhou2009ionic}. Another significant advantage of ionic liquids is the large variety of cations and anions that can be used: the chemistry of ionic liquid synthesis makes it easy to ``mix-and-match'' cations and anions. This property means that ionic liquids can be tuned for a particular application, with the cations and anions playing their own roles such as adsorbing onto the surfaces to protect them from wear-and-tear \cite{qu2009tribological, somers2013review,xiao2016ionic}. 

The chemistry of ionic liquids aside, a fundamental property of ionic liquids is that they are ionic, i.e. there are ions in the fluid that can respond to an applied electric field. As such, it is natural to wonder whether the ionic nature of ionic liquids or electrolytes can be exploited to control friction using electric fields. For ionic liquids, seminal works by Perkin \emph{et al.} show that nanoconfined ionic liquids near a charged surface arrange themselves in a layered structure of alternating cation-rich and anion-rich layers \cite{perkin2010layering,perkin2011self}. The number of ion layers confined between the surfaces is an integer dependent on the surface separation, and thus the friction coefficient in the low velocity regime also shows a ``quantized'' behaviour as a function of surfaces separation \cite{smith2013quantized}. Applying a potential difference between the surfaces and the bulk switches the composition of the ion layers between the surfaces from cation-enriched at negative potentials to anion-enriched at positive potentials. This offers a handle to tune the friction coefficient if the lubricating properties of the cations and anions are different \cite{sweeney2012control,li2014ionic}. On the theory front, molecular dynamics simulations corroborated the importance of ion layering in determining friction response and attributed electric field effects to structural changes in the ion layers \cite{capozza2015electrical,fajardo2015electrotunable,fajardo2015electrotunable1}. Nonetheless, the aforementioned theories and simulations focus on the regime of low sliding velocity and nanoconfined ionic liquids, which is perhaps less relevant for industrial applications such as lubrication in engines. Moreover, in the nanoconfined regime, molecular parameters such as the ion size and shape enter into the problem in addition to the fundamental physics of ion-ion electrostatic interactions.  

To isolate the effect of ion-ion electrostatic interactions on friction, a simpler system to consider is dilute electrolytes. For micron-scale surface separations and moderate sliding velocities that are still below the transition to turbulence, it is known that charged surfaces experience a larger friction in dilute electrolytes compared to uncharged surfaces \cite{bo1998hydrodynamic,benard2009enhanced}. This is because ions arrange themselves near an oppositely charged surface forming an electrical double layer, and the electrical double layer opposes flow which disrupts its structure. Qualitatively, the ion layer near the surface ``holds on'' to the fluid, and thus decreases the effective separation between the surfaces and therefore increases hydrodynamic friction. However, drag reduction is accomplished in a related system of dielectric barrier discharge plasma actuators: gas molecules are ionized around a object in a flow (e.g. an airplane wing) to create a plasma. Electrodes on the object manipulate the electrical double layer between the plasma and the electrodes to reduce drag by generating a directional plasma body force \cite{corke2010dielectric}. This suggests that electrolyte systems are fruitful systems to explore drag reduction with active flow control. 

Nonetheless, at industrially relevant sliding velocities, we would expect the fluid flow to be turbulent. In order to achieve drag reduction in a turbulent flow, what is needed are mechanisms which modify the turbulence production cycle, more specifically the transfer of shear stress between flow and wall through turbulent structure (the so-called ``momentum cascade'') \cite{jim12}. This is because the production of near-wall turbulence and friction drag are intimately linked \cite{tow76}. Modifying and stabilizing this cycle has been shown to be possible using deformable bubbles \cite{lu05,ber05}, large polymer chains \cite{lum69} and small grooves or riblets \cite{cho93}. More recently, Du and Karniadakis \cite{du00} showed it is theoretically possible to reduce drag using traveling wave forces. All of these mechanisms modify the near-wall vortical structures, also known as streaks, and restrict their production and interaction, resulting in large decrease in drag. However, the coupling between the turbulence production cycle and the electrical double layer in an electrolyte is, to our knowledge, unexplored.  

In this paper, we will address the question of whether proposals of friction control in the slow flow regime by applying an electric field across an electrolyte lubricant can also be successful in the turbulent regime. The model system we will focus on is a turbulent plane Couette flow, i.e. the shear flow between two parallel plates.  To do this, we will first present a computationally efficient algorithm for solving the coupled Poisson-Nernst-Planck Navier-Stokes equation. We will then study the effect of an applied electric field on a turbulent plane Couette flow of electrolyte. Our results show that the effect of a static electric field on drag in the turbulent regime is minimal in the parameter regime simulated. 

\section{Numerical methods} 

To model a dilute electrolyte, the incompressible Navier-Stokes equations with an electric body-force:

\begin{equation}
 \rho_w \left ( \displaystyle\frac{\partial \textbf{u}}{\partial t} + \textbf{u}\cdot\nabla\textbf{u} \right )=-\nabla p + \eta \nabla^2 \textbf{u} + \rho \nabla \Phi,
 \label{NS_eqn}
\end{equation}

\begin{equation}
\nabla \cdot \textbf{u}=0,
\label{incompressibility}
\end{equation}

\noindent are coupled with the Poisson-Nernst-Planck (PNP) equations for positive and negative ion concentration fields:

\begin{equation}
 \displaystyle\frac{\partial c_\pm}{\partial t} + \textbf{u}\cdot\nabla c_\pm =D \nabla \cdot ( \nabla c_\pm \pm V_T^{-1} c_\pm \nabla \Phi),
 \label{eq:PNP}
\end{equation}

\noindent and Poisson's equation for the electric potential:

\begin{equation}
 -\epsilon \nabla^2 \Phi = \rho,
  \label{eq:potential}
 \end{equation}

\noindent where $\textbf{u} \equiv (u_x,u_y,u_z)$, $p$, $\rho_w$ and $\eta$ are the fluid velocity ($u_i$ being the $i^{th}$ spatial component), pressure, density and dynamic viscosity respectively, $t$ is time, $c_\pm$ is the positive (negative) ion concentration, $\rho = e(c_+-c_-)$ the charge density, $e$ the charge on an electron, $\Phi$ the electric potential, $D$ the ion mobility, $V_T = k_B T/e$ the thermal voltage, $k_B$ the Boltzmann constant, $T$ the fluid temperature and $\epsilon$ the dielectric permittivity. We assume that the cations and anions are univalent. Equation (\ref{NS_eqn}) describes momentum conservation for the fluid, with the forcing due to ion transport, $\rho \nabla \Phi$, entering into the momentum balance; the physical assumption that the fluid is incompressible is imposed by Equation (\ref{incompressibility}). The Poisson-Nernst-Planck system, Equations (\ref{eq:PNP})-(\ref{eq:potential}), is a set of coupled convection-diffusion equation describing the transport of ions due to migration in response to an electric field, diffusion in the solvent, and advection by the background flow. The electric field is, in turn, computed self-consistently from the charge density via the Poisson equation (\ref{eq:potential}). The stationary solution of Equations (\ref{NS_eqn})-(\ref{eq:potential}) is the Poisson-Boltzmann equation. We note that Equations (\ref{NS_eqn})-(\ref{eq:potential}) are only a minimal model for electrolyte transport -- steric effects as well as beyond mean field ion-ion correlations are neglected in Equations (\ref{NS_eqn})-(\ref{eq:potential}). We will comment on those effects in a later section of this paper. 

The geometry we consider is the flow between two parallel and infinite plates (electrodes) separated a distance $2h$, and moving in opposite directions with velocities $\pm U$, which is commonly referred in fluid dynamics as a plane Couette flow. We take the $y$ direction as the wall-normal direction, the $x$ direction as the stream-wise direction (direction parallel to the flow) and the $z$ direction as the span-wise direction. The infinite $x$ and $z$ directions are modeled as periodic, with periods of $L_x$ and $L_z$ respectively. The velocity boundary conditions are no-slip and no-penetration at the electrodes, while the ion boundary conditions are no-penetration and a fixed electric potential. For this manuscript, we consider a potential of $\pm V$ at the plates, constant in space and time, but the code accepts any potential at the boundaries.

As a base, we use the AFiD code to solve the incompressible Navier-Stokes equations. AFiD uses a centered second order finite difference scheme on a staggered mesh for spatial discretization, with a third-order low-storage Runge-Kutta fractional-step time marching for the non-linear terms and a second order Adams-Bashforth scheme for the viscous and pressure terms. The incompressibility condition is enforced through a projection-correction method. To prevent having to solve large sparse matrices, a tridiagonal approximate factorization is used. The code has been parallelized using MPI directives and has ran on up to 64,000 cores. More details about the Navier-Stokes solver, validation procedures and performance can be found in Refs. \cite{verzicco,poel15}. 

\begin{figure}
\centering
\hspace{1cm}
\begin{tikzpicture}
\draw (0,0) rectangle (4,4);
\filldraw[black] (2,2) circle(1mm);
\node [above right,black] at (2,2) {$c_\pm,\Phi$};
\filldraw[black] (0,2) circle(1mm);
\node [above right,black] at (0,2) {$j_x$};
\filldraw[black] (4,2) circle(1mm);
\node [above right,black] at (4,2) {$j_x$};
\filldraw[black] (2,0) circle(1mm);
\node [above right,black] at (2,0) {$j_y$};
\filldraw[black] (2,4) circle(1mm);
\node [above right,black] at (2,4) {$j_y$};
\draw [->,thick] (4.5,0) -- (5.5,0);
\draw [->,thick] (4.5,0) -- (4.5,1.0);
\draw [->,thin, dashed] (4.5,0) -- (5.2,0.7);
\node [above,black] at (4.5,1.0) {$y$};
\node [right,black] at (5.5,0) {$x$};
\node [above right,black] at (5.2,0.7) {$z$};
\end{tikzpicture}
\caption{Location of electric potential, concentrations and ion fluxes of a 2D simulation cell. The third dimension ($z$) is omitted for clarity. The electric potential and the concentrations are placed in the cell center, but the ion fluxes are placed on the borders of the cells (the same place as the velocities), to ensure ion conservation at the walls. This allows the code to enforce the boundary condition $j_y=0$ to machine precision at the walls.}
\label{fig:disc}
\end{figure}
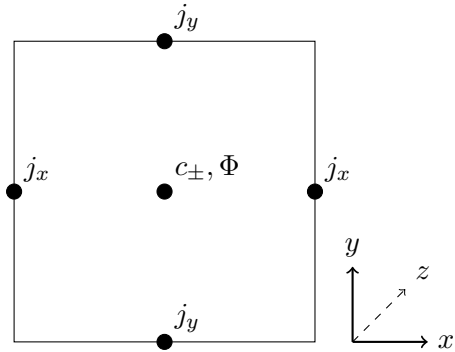

The PNP equations (\ref{eq:PNP}) require special attention when discretizing. An advection-diffusion equation for a scalar field, even when dealing with an active scalar such as the temperature field in thermal convection discussed in Ref. \cite{poel15}, can be discretized in rather na\"ive ways and still produce somewhat accurate results. But ion-concentration fields have exponential boundary layers (the electrical double layer) which can cause problems with ion conservation if not treated correctly. Discretization errors of the order of second and third derivatives can be quite high close to the walls. The electrical double layer makes the equation very stiff, and ideally the terms should be treated fully implicitly. Indeed, we solve the diffusive term ($\nabla^2 c_\pm$) in an implicit way, analogous to the viscous terms in the momentum equation. However, the ion flux due to migration in response to the electric field ($\textbf{j}_\pm=\pm D V_T^{-1} c_\pm \nabla \Phi $), is a non-linear term, so the implicit solution is more complicated. Instead, we stagger $\textbf{j}$ on the cell boundaries, so as to enforce the no-penetration condition at the electrodes exactly and solve the term explicitly. Figure \ref{fig:disc} shows the staggered arrangement, the $z$ (spanwise) direction is omitted for clarity.

To solve the electric potential, we decompose it into a base potential $\Phi_0$, which satisfies the boundary conditions, and a perturbation $\phi$, which equals to zero at the boundaries. In our case, we simply take $\Phi_0 = Vy/h$, and solve $\nabla^2 \phi = -\rho/\epsilon$ using an exact Poisson solver. In the same way as the pressure correction step is implemented (see  Refs. \cite{verzicco,poel15} for an extended discussion), we Fourier transform the two homogeneous directions and solve the resulting tridiagonal matrix using the Thomas algorithm in the wall-normal direction. The only difference is that the pressure field has a Neumann boundary conditions ($\partial_n p=0$) at the wall, while the potential perturbation has a Dirichlet boundary condition ($\phi=0$) at the wall.

The resulting system has two geometrical parameters, the periodicity ratios in the span-wise $\Gamma_x = L_x/h$ and stream-wise $\Gamma_z = L_z/h$ directions, and five dimensionless control parameters. Three control parameters are properties of the electrolyte. First, we have the Debye length relative to the plate separation 
\begin{equation}
\lambda_D/h=\sqrt{\epsilon V_T/(2c_0 e)}/h,
\end{equation}
which is a measure of characteristic width of the electrical double layer near the electrodes. Second, the ratio between the electrical forces and the viscous forces in the electrolyte is given by the dielectric coupling constant 
\begin{equation}
 \beta=\epsilon V_T^2/\eta D,
 \label{eq:kappa}
\end{equation}
which is usually taken to be $\beta=0.5$ for dilute electrolytes. Third, the ratio between momentum diffusivity to ion mobility is given by the Schmidt number
\begin{equation}
 Sc=\eta/\rho_wD. 
\end{equation} 
For a typical electrolyte solution such as salty water, $Sc \approx 1000$. Computational challenges are associated to high $Sc$ simulations as conservative estimates based on dimensional analysis imply that the ion field would require  $\mathcal{O}(Sc^{1/2})$ more grid points in every direction (c.f. Ref \cite{multires} for a full discussion). Thus an estimate for the resolution required for $Sc=1000$ would be $Sc^{3/2}\sim 30000$ more points to properly resolve the ion field. Here we limit ourselves to cases with Schmidt numbers close to unity to make the simulations possible. We would expect that this approximation does not significantly affect the physics as $Sc$ is a measure of ion diffusion. Due to the intense turbulent mixing, the motion of ions is dominated by the chaotic turbulent mixing (coupled to the ion dynamics through the advection term) and the diffusive term in the PNP equations is comparatively less significant, thus the approximation is likely valid. A detailed study on the effects of $Sc$ is left for future work.

The remaining two control parameters are the dimensionless voltage
\begin{equation}
 \hat{V}=2V/V_T,
\end{equation}
and the Reynolds number
\begin{equation}
 Re=\rho_w U h/\eta.
\end{equation}
The dimensionless voltage measures of the electrical forcing of the system. In the absence of flow, once $\hat{V}$ exceeds around unity, large gradients of concentration develop near the electrodes, i.e. the electrical double layers. Here we will focus on values of $\hat{V}$ between $2$ and $40$. The Reynolds number is a measure of the strength of the shear driving of the fluid flow compared to the viscous forces. This control parameter is independent of the electrokinetics, and will be fixed to $Re=3000$ which is a mildly turbulent plane Couette flow benchmarked in the literature \cite{pirozolli}. 

For convenience, we also define the following dimensionless quantities: the frictional force $F$ at the wall is simply the average shear stress $\tau_w=\eta\partial_y \langle u_x\rangle $  multiplied by the plate area $A$; the average is computed by averaging over time as turbulence is a chaotic phenomenon. Friction can be non-dimensionalized either as a friction coefficient $c_f = 2\tau_w/(\rho_w U^2)$ , or a frictional Reynolds number $Re_\tau = \rho_W u_\tau h/\eta$, where $u_\tau = \sqrt{\tau_w/\rho}$. We define the viscous length as $\delta_\nu = \eta / (u_\tau \rho_W)$. The viscous length is analogous to the Debye length, and gives us an estimate of the minimum length scale for momentum structures, below which they are smeared out by viscosity; near-wall structures are of minimum size $\sim 10 \delta_\nu$ \cite{avskariskov}. The ratio $d/\delta_\nu$ is simply $Re_\tau$, so comparing $Re_\tau$ and $h/\lambda_D$ gives us an idea of the locations of the momentum and electrokinetic structures.

To validate the code we run two test cases. One with the flow completely decoupled from the electric field to test the fluid mechanics, and one with steady walls to decouple the dynamics of ion transport from the background turbulent mixing. For the hydrodynamic test, we set $\Gamma_x=2\pi$ and $\Gamma_z=\pi$ for the simulation. We set $Re=3000$, a well-used validation case and use a grid resolution of $256\times512\times256$ points uniformly distributed in the horizontal directions, and clustered near the wall in the wall-normal direction. The simulation is ran until a stationary state is achieved, and statistics are then taken for $200$ time units based on $d/U$.  We obtain $c_f=5.63\cdot 10^{-3}$ and $Re_\tau = 172$, in line with the results of \cite{pirozolli}.  

To validate the electrokinetic module, a one-dimensional simulation with a quiescent background flow (achieved by setting the wall velocities equal to zero) is performed. The relevant control parameters are set to $\hat{V}=4$ and $\lambda_D/h=100$. A grid resolution of a total of 512 points in the wall-normal direction is used, with more grid points positioned near the walls. The simulation is started from a homogeneous concentration field and marched in time until the concentration profiles have reached a stationary state. The value of $Sc$ does not play a role, as it just gives a time-scale for equilibration. Here we set it to unity. The other parameter, $\beta$, also has no role as there is no coupling between fluid momentum and electric field. 

\begin{figure}[!h]
\centering
  \includegraphics[width=0.49\textwidth]{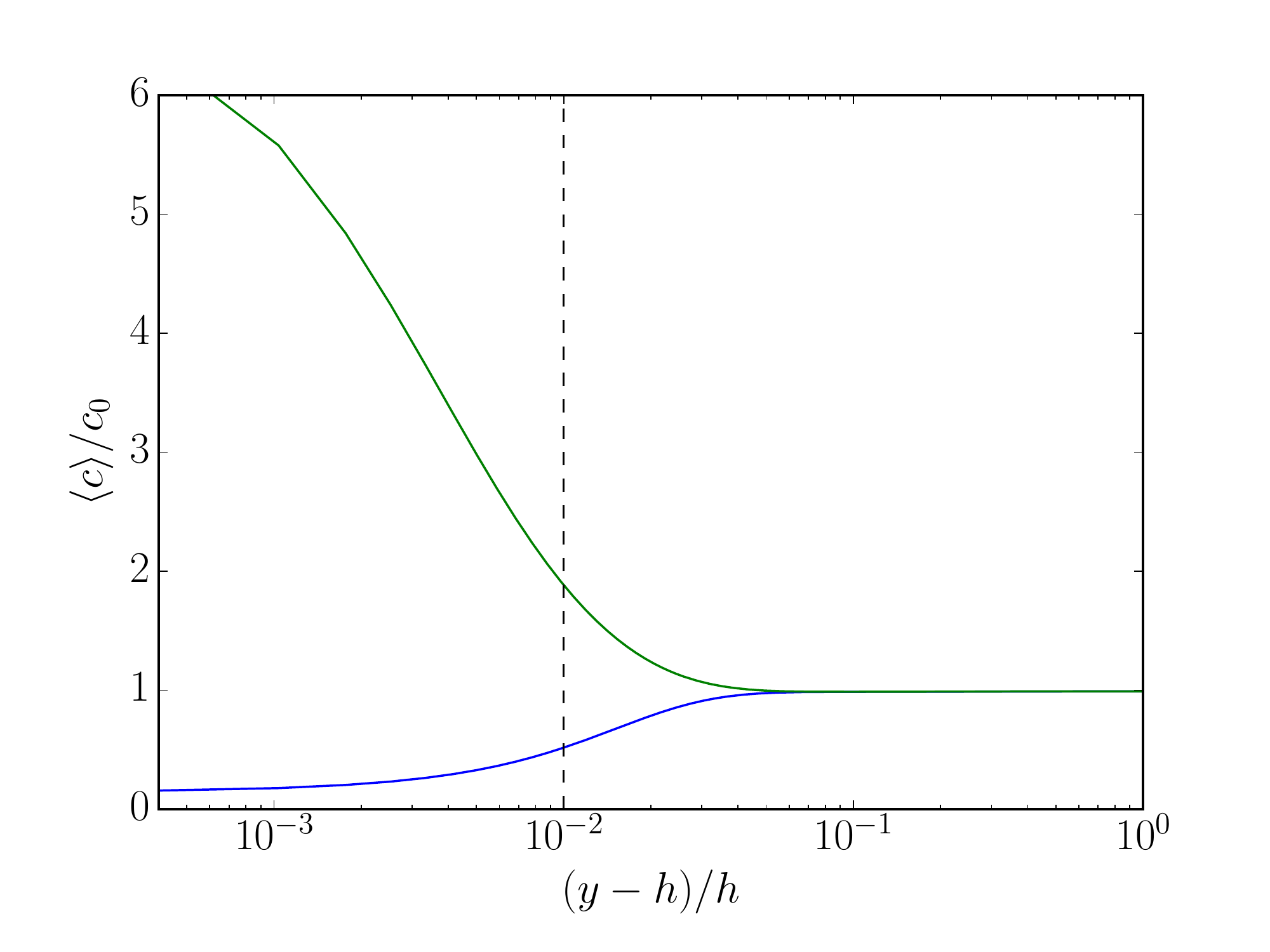}%
    \includegraphics[width=0.49\textwidth]{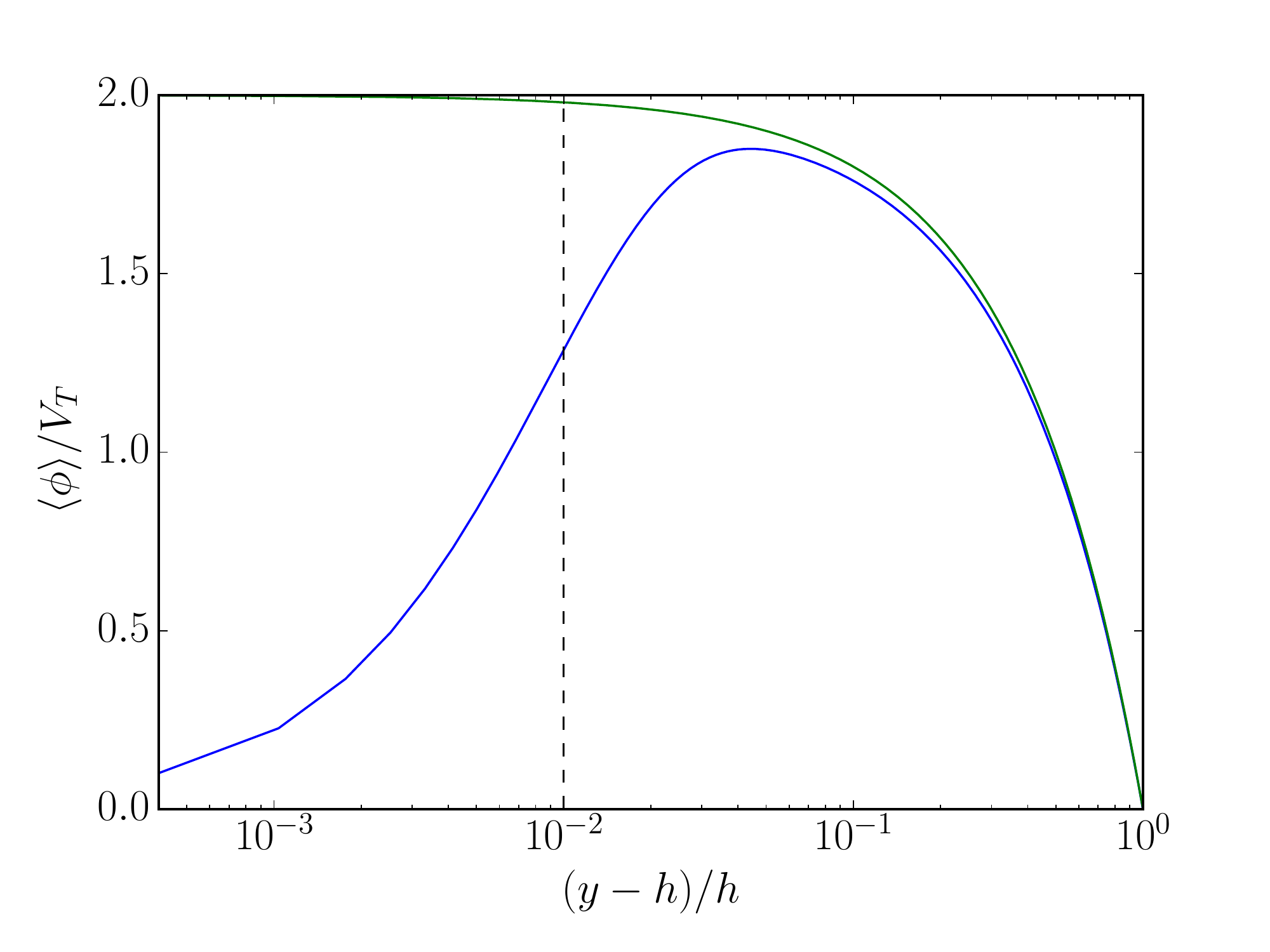}%
\caption{Left: Average positive (blue) and negative (green) ion concentration profiles for no flow, $h/\lambda_D=100$ and $\hat{V}=4$. Right: Resulting electric potential perturbation (blue) and electrical potential perturbation required for an electrically neutral bulk (green). The vertical dashed line on both plots indicates the position of a point one Debye length away from the wall. }
\label{fig:validation}%
\end{figure}

The resulting average concentration fields are shown in the left panel of Figure \ref{fig:validation}. The concentration field decays exponentially away from the electrode with a characteristic decay length of the order of the Debye length. The right panel shows the resulting electric potential perturbation and the baseline solution $\phi=-\Phi_0=-Vy/h$ in green. The electric field perturbation approaches this line in a few Debye lengths. Furthermore, the profiles are symmetric respect to the centre, and the total ion quantity is conserved. 
 
\section{Results}

With the previous tests, we are now confident our code produces valid results. We numerically investigate three sets of parameters by varying $\hat{V}$ between $2$ and $40$. The full set of dimensionless parameters for the three cases are tabulated in Table \ref{tbl:numdet}. The grid resolution used is the same as for the validation cases, i.e. $256\times512\times256$. As the $Sc$ number has been fixed to $3$, there are two further parameter choices --- varying the electrokinetic coupling constant $\beta$ fixing $h/\lambda_D$ (L100 and L100B cases shown in Table \ref{tbl:numdet}) and varying $h/\lambda_D$ fixing $\beta$ (L100 and L33 cases shown in Table \ref{tbl:numdet}). 

\begin{table}[!h]
\centering
\begin{tabular}{cccc}
Case Name & L100 & L100B & L33 \\
$Re$ & 3000 & 3000 & 3000 \\
$Sc$ & 3 & 3 & 3 \\\
$h/\lambda_d$ & 100 & 100 & 33 \\
$\beta$ & 0.5 & $1.5\cdot10^{-4}$ & 0.5 \\
\end{tabular}
\caption{\label{tbl:numdet}Control parameters for simulation cases.}
\end{table}

\begin{figure}[!h]
\centering
  \includegraphics[width=0.60\textwidth]{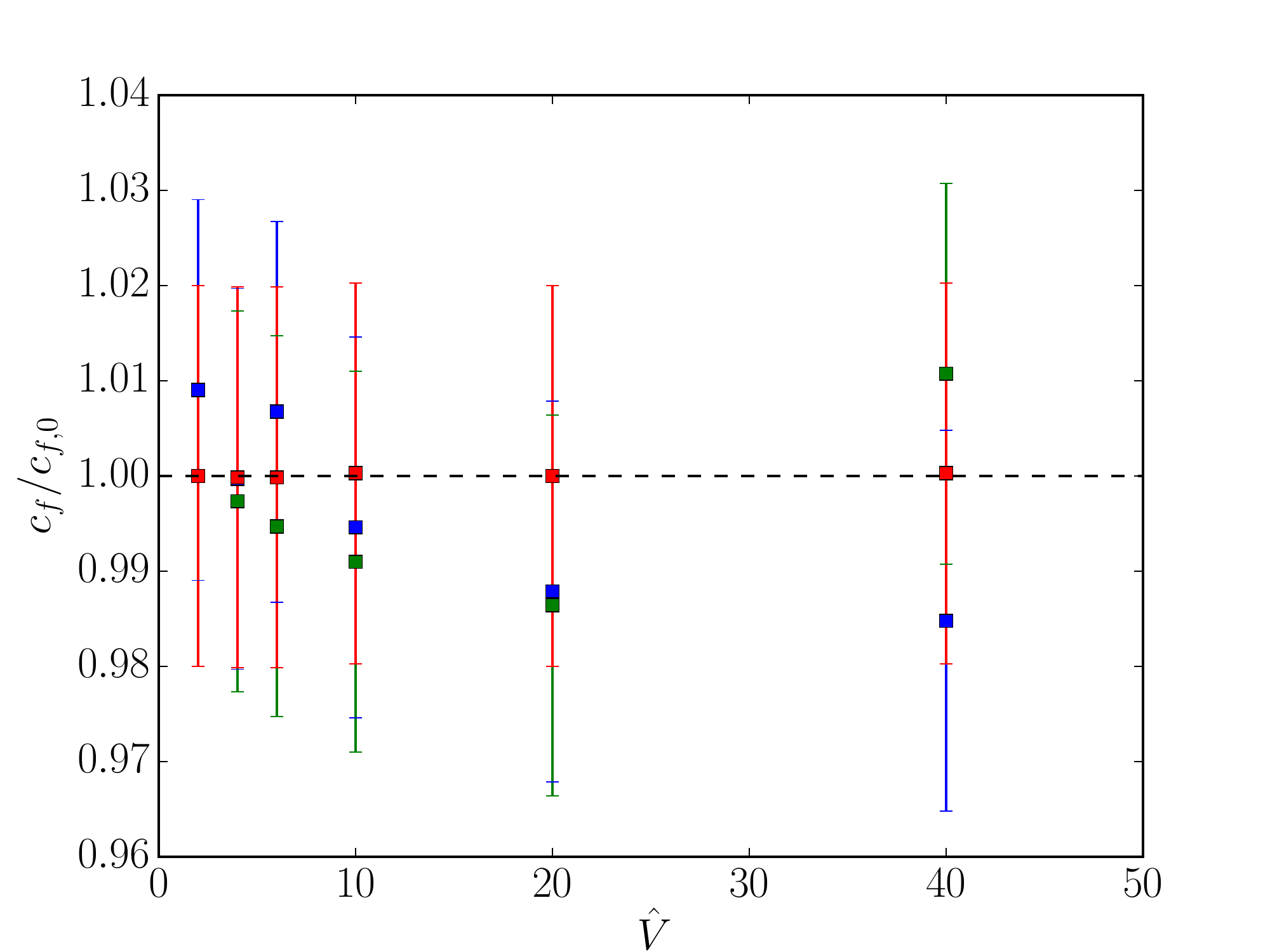}%
\caption{Friction coefficient normalized with the baseline case $c_{f,0}$ against the dimensionless voltage $\hat{V}$ for the three sets of simulations. Symbols: red squares L100B, green squares L100, and blue squares L33.}
\label{fig:results}%
\end{figure}

Figure \ref{fig:results} shows the friction coefficient against the dimensionless voltage for the three sets of simulations. No discernible effect on the drag outside the error bars can be seen for all cases. The effect of an electrical double layer appears negligible. While a small trend does seem to appear for the L33 case, this trend is inside the error bars, and, even if significant only results in a small drag reduction of $1-2\%$ at $\hat{V}=40$. The L100 case also shows a change at $\hat{V}=40$, which could be caused by the fact that the flow perturbations coming from the electrical double layer become significant. However, comparing the mean velocity profiles and the mean concentration profiles to the baseline cases with no flow or no electric field show no significant interaction between the ions and the mean flow in the L100 case. Finally the L100B case shows almost no deviations from the baseline case, highlighting the importance of $\beta$, the ratio between the electrical forces and the viscous forces, as a control parameter. 

\begin{figure}
\centering
  \includegraphics[width=0.49\textwidth]{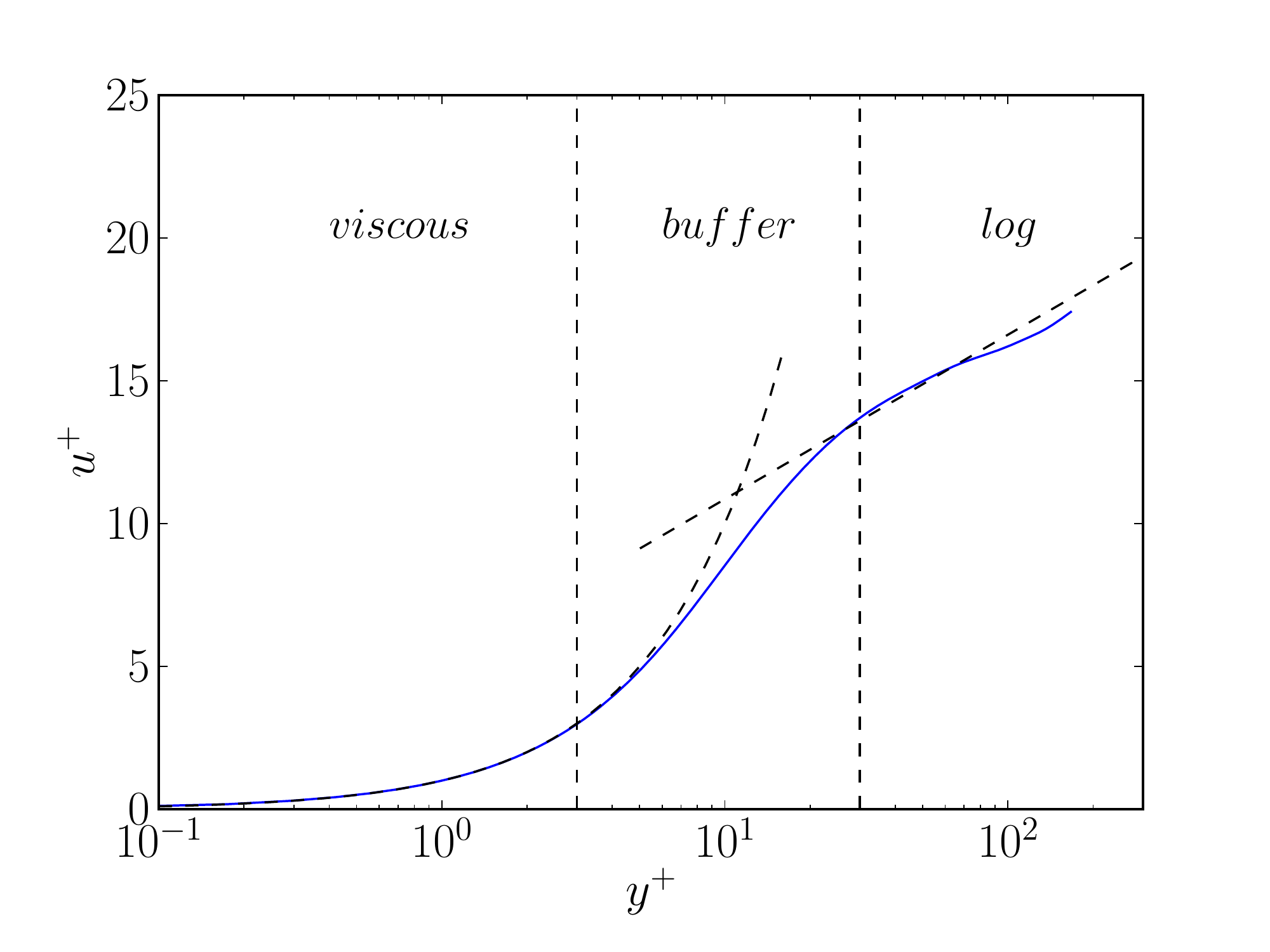}
    \includegraphics[width=0.49\textwidth]{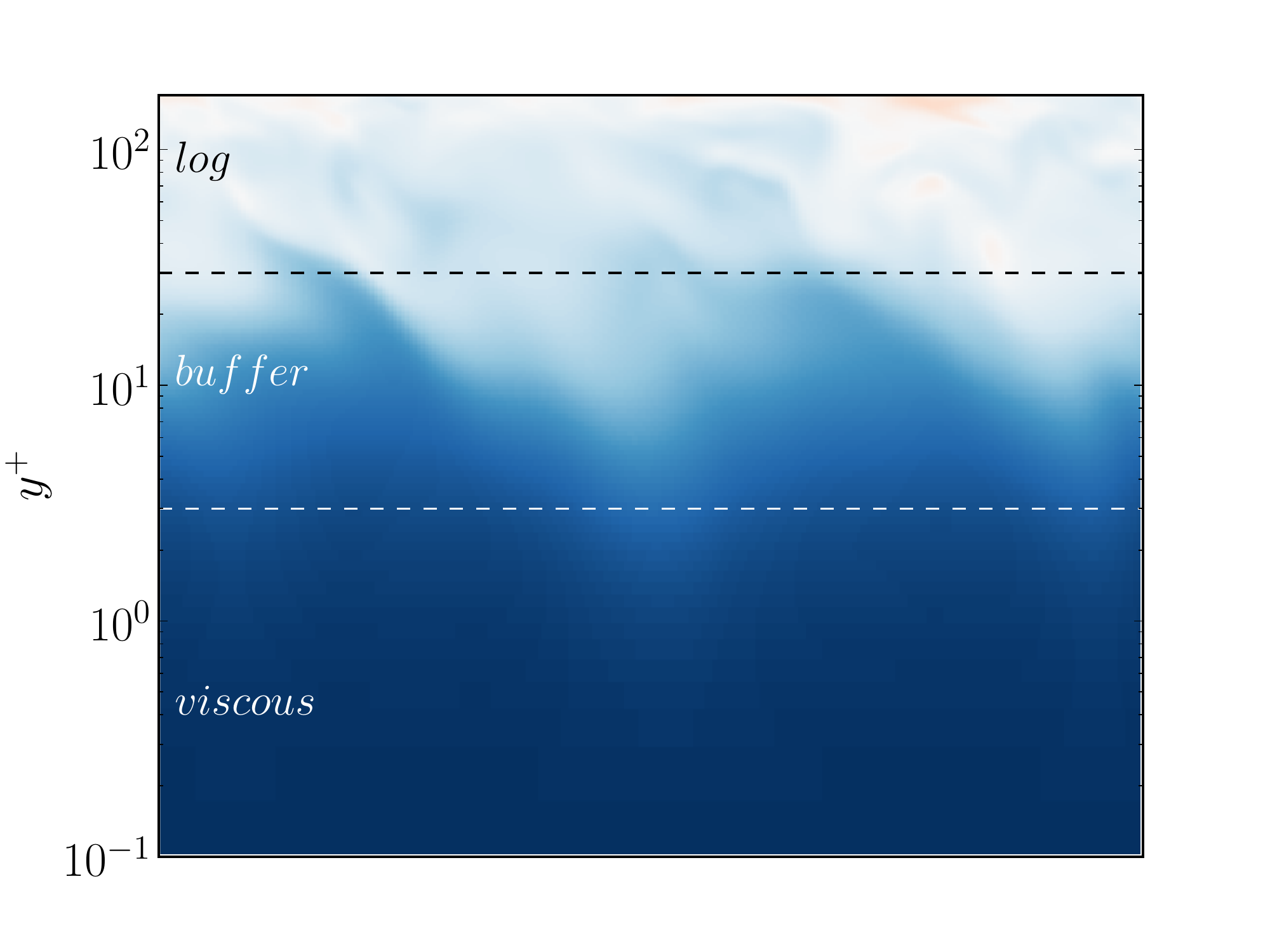}%
\caption{Left: Mean streamwise velocity against wall distance in wall units, and sub-division of turbulent boundary layer. Dashed lines indicate $u^+=y^+$ and $y^+=2.5\log(y^+) + 5.2$, the theoretical results for the viscid sublayers and logarithmic layers. Right:  visualization of instantaneous streamwise velocity across the three layers. The viscid sub-layer is laminar, with barely no streaks, i.e. turbulent structures. The buffer sub-layer shows some streaks, and the logarithmic sub-layer is fully dominated by turbulence. Red and blue indicate the two velocity extrema, those of the walls, and a paler colour indicates a lower flow speed. }
\label{fig:validationvv}%
\end{figure}

To understand why the effect of the electrical double layer on friction is minimal, we briefly sketch out the structure of the turbulent flow near the wall. Figure \ref{fig:validationvv} shows the mean stream-wise velocity profiles in wall-units, that is, the wall-normal coordinate normalized by the viscous length $y^+=y/\delta_\nu$, and the mean-streamwise velocity deficit $u^+=(U-u)/u_\tau$ for the validation case discussed in the previous section. These wall units are chosen as the boundary layer velocity profiles are close to universal when plotted in those units. 

Three regions (or sub-layers) can be distinguished in the plot: the viscous, buffer and logarithmic sub-layers. The viscid sublayer is the closest to the wall, and has the velocity profile $y^+=u^+$ (shown in the figure). In this sub-layer, which extends up to $y^+\approx 5$, little to no turbulence is present and the effect of viscosity dominates. When $y^+$ is larger than $30$, there exists a layer called the logarithmic layer, as it follows the empirically found law $u^+=2.5\log(y^+)+5.2$. In our case, $Re_\tau\approx170$, thus this region is still developing and deviations from the logarithmic profile are expected. The logarithmic sub-layer is fully turbulent, and is dominated by the interaction of turbulent vortical structures such as streaks. 

Between both regions is the buffer sub-layer. The buffer sub-layer contains the peak production of turbulent energy at $y^+\approx 13$. This peak production region is the region which we want to affect in order to break the physical processes which form turbulence. Breaking the turbulence generation process would in turn reduce drag. The values of $h/\lambda_d$ in L100, L100B and L33 are chosen such that the viscous wall-unit and the Debye length scale are of comparable size. Therefore with these parameter values we would expect the electrical double layer to lie near the buffer region of the boundary layer, thus the fluctuations of concentration fields induced by the turbulence could significantly couple back into the flow.  For the electrical perturbation to couple back to the flow, the inertial forces ($\rho_W \textbf{u} \cdot \nabla \textbf{u}$) and the electrical forces ($\rho_e \nabla \phi$) must be of comparable strength. For this to happen, we expect the condition:

\begin{equation}
 \beta Sc^{-1} \hat{V}^2 \gtrsim \mathcal{O}(1)
 \label{eq:one}
\end{equation}

\noindent to be satisfied, as long as $\lambda_d$ and $\delta_\nu$ are of similar magnitudes (see Appendix for a derivation). As we have lowered $Sc$ to ensure numerical stability, we expect to see an effect even at the lower end of the $\hat{V}$ range. 

\begin{figure}
\centering
    \includegraphics[width=0.60\textwidth,trim={0 3cm 0 3cm},clip]{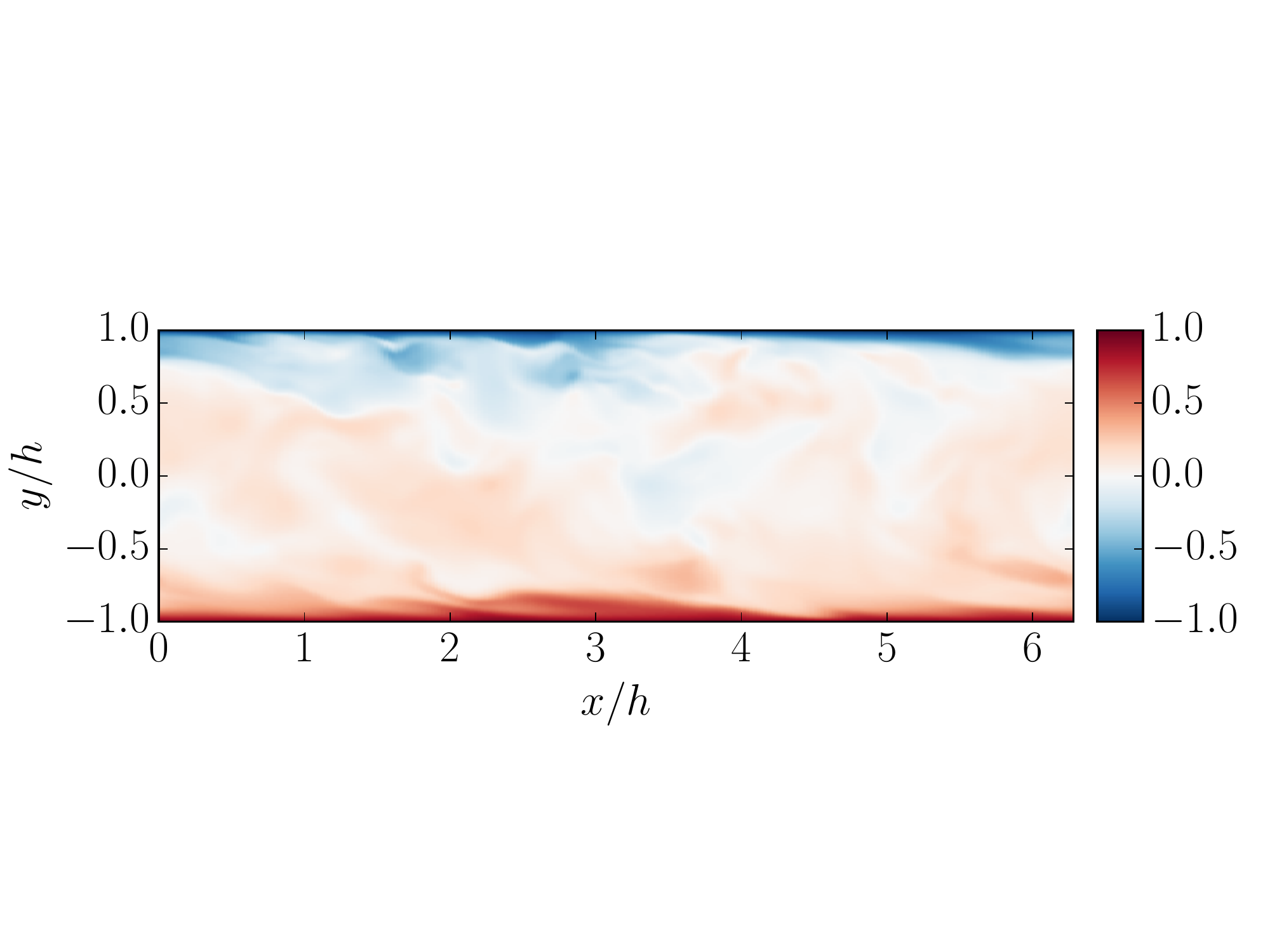}\\%
  \includegraphics[width=0.60\textwidth,trim={0 3cm 0 3cm},clip]{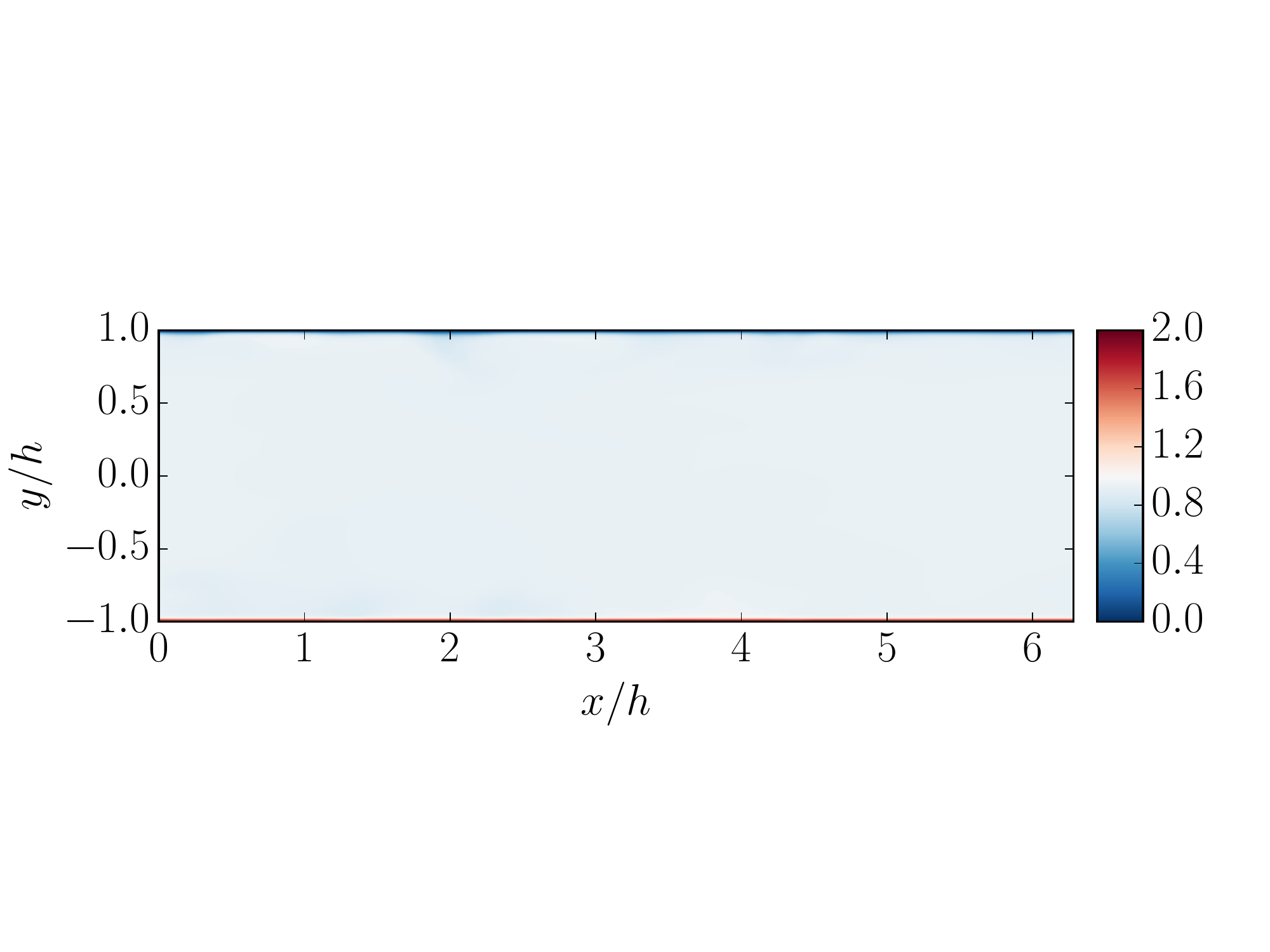}%
\caption{Top: visualization of a streamwise cut of the instantaneous streamwise velocity for the L100 case and $\hat{V}=10$. Red and blue indicate the two velocity extrema, those of the walls, and a paler colour indicates a slower flow. Bottom: Same visualization, now for the $c_+$ field. Red indicates high concentration (located near the bottom electrode), blue indicates low concentration (near the top electrode), and a pale colour indicates the mean concentration $c_0$.}
\label{fig:visuals}%
\end{figure}

Figure \ref{fig:visuals} shows that the effect of the ions density on the flow is indeed localized near the walls. The top panel of Figure \ref{fig:visuals} shows the instantaneous streamwise velocity for the L100 case at $\hat{V}=10$. The characteristic bending of the turbulent streaks in the direction of the flow can be seen. The bottom panel of Figure \ref{fig:visuals} shows the instantaneous positive ion concentration field.  Barely any perturbation to the ion density can be seen in the bulk.

\begin{figure}
\centering
  \includegraphics[width=0.49\textwidth]{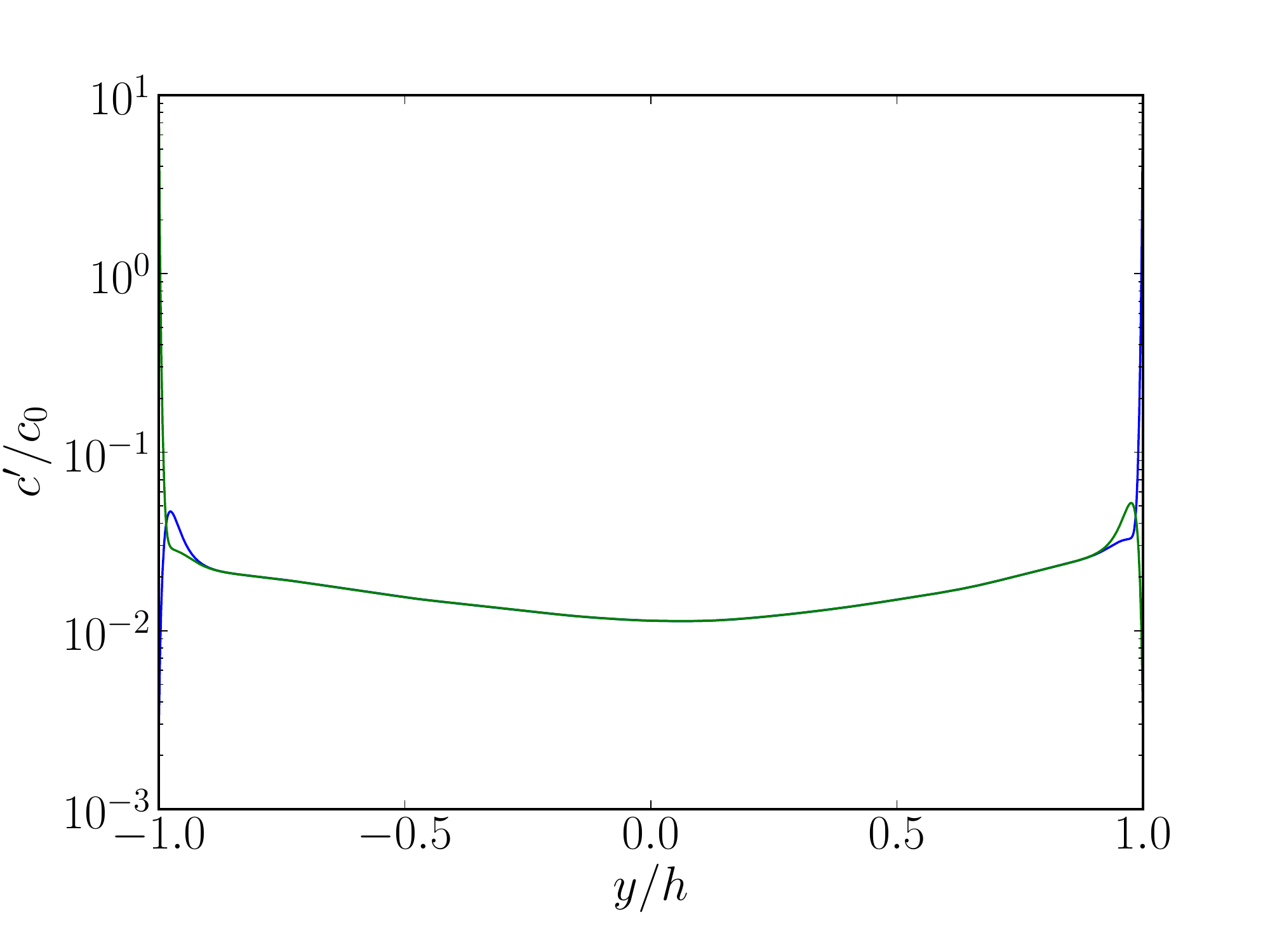}%
  \includegraphics[width=0.49\textwidth]{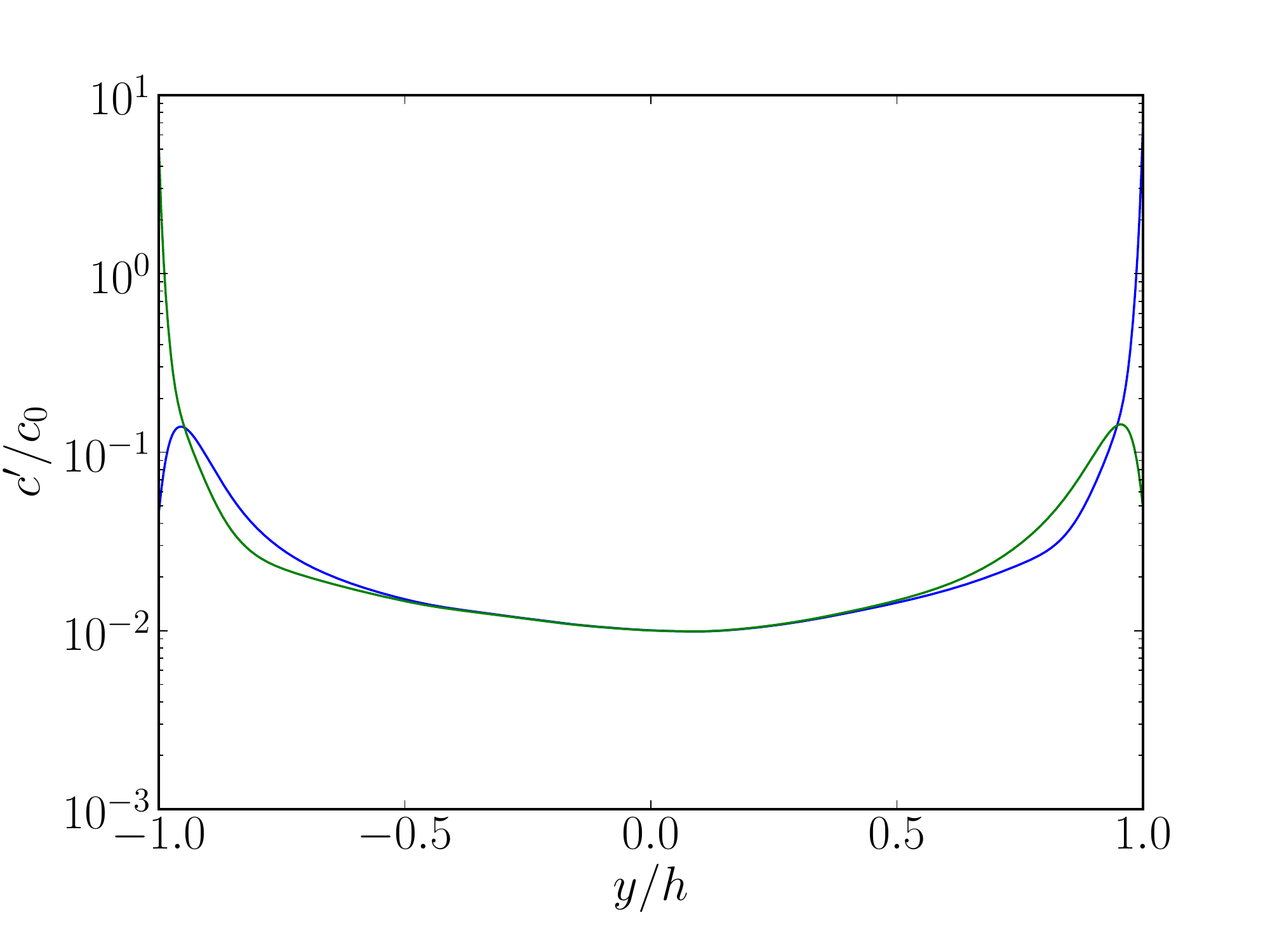}%
\caption{RMS fluctuations for the ion concentration fields for the L100 (left) and L33 (right) cases with $\hat{V}=10$. Blue is $c_-$ and green is $c_+$.}
\label{fig:results1}%
\end{figure}

To further understand the effect of the electrical double layer on fluid flow, we show in Figure \ref{fig:results1} the root mean squared fluctuations of the ion fields for the L100 (left) and the L33 (right) cases with $\hat{V}=10$. For both panels, we can again see that the ion fluctuations barely penetrate into the bulk. Fluctuations appear near the wall, inside the turbulent boundary layer, where the mechanisms for production of turbulence (and drag increase) are located. For the L100 (and L100B) case, the fluctuation peak inside the boundary layer appears at around $4$ wall-units from the electrodes, inside the viscid sublayer of the turbulent boundary layer. In the viscid sublayer, there is little turbulence, so the effect the electrical double layer on turbulence is minimal. However, for the L33 case, the peak in fluctuations appears at $10$ wall units. In the turbulence generation cycle, this is very close to the area where most turbulent energy is produced (c.f. Figure \ref{fig:validationvv}), and also the peak in the fluctuations is larger. Thus the L33 case has a higher chance of affecting the structure of turbulence generation and drag increase; this is indeed inline with the results in Figure \ref{fig:results}, although the trend cannot be conclusively ascertained within numerical errors. 

\begin{figure}
\centering
  \includegraphics[width=0.49\textwidth]{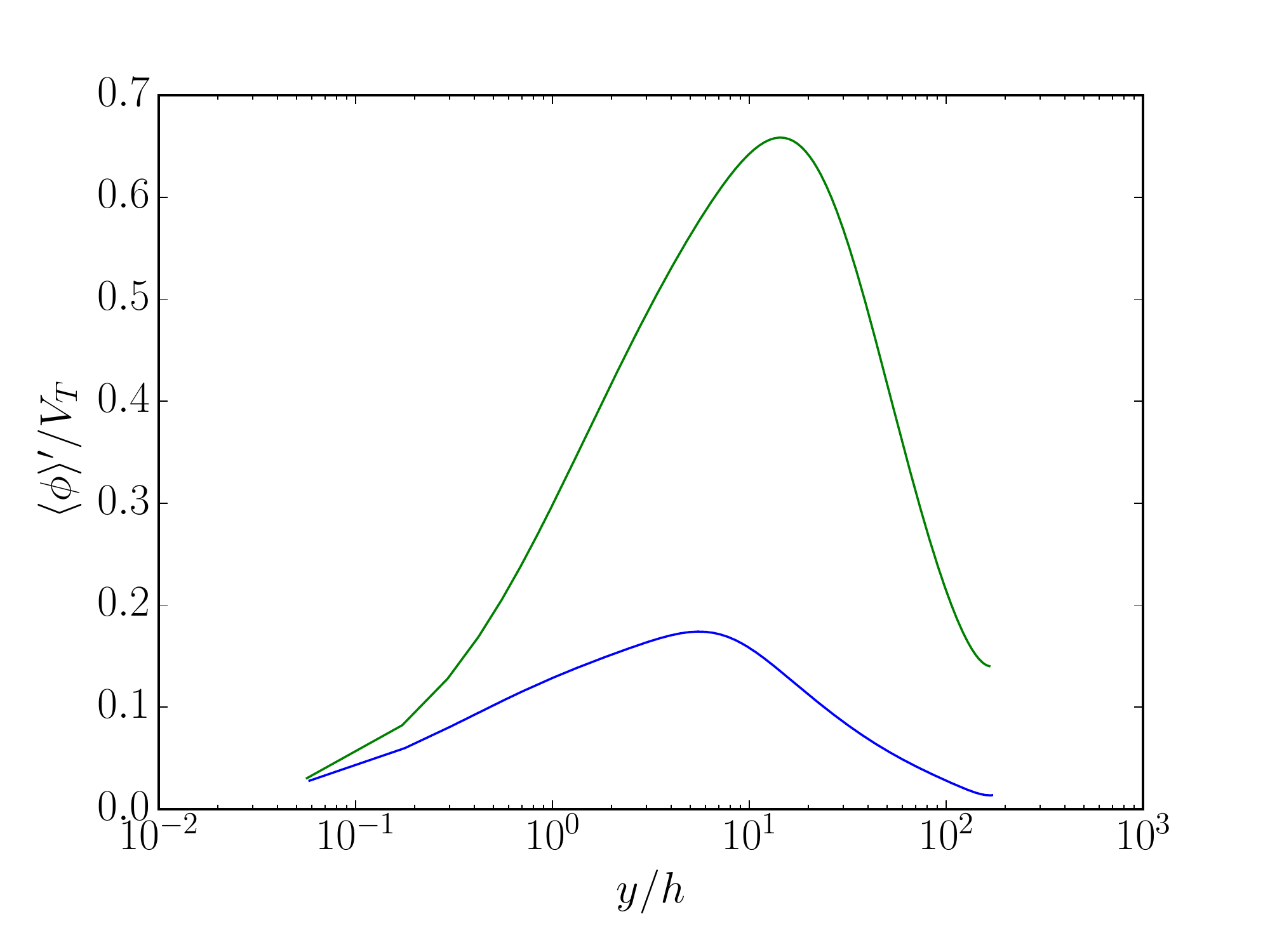}%
\caption{RMS fluctuations for the electric potential fields for the L100 (blue) and L33 (green) cases with $\hat{V}=10$. The peak in $\phi$ fluctuations is much higher when it coincides with the peak turbulent energy production at $y^+=13$. The profiles are shown up to mid-gap, both walls are symmetrical. }
\label{fig:rmsphi}%
\end{figure}

\begin{figure}
\centering
  \includegraphics[width=0.45\textwidth,trim={0 3cm 0 3cm},clip]{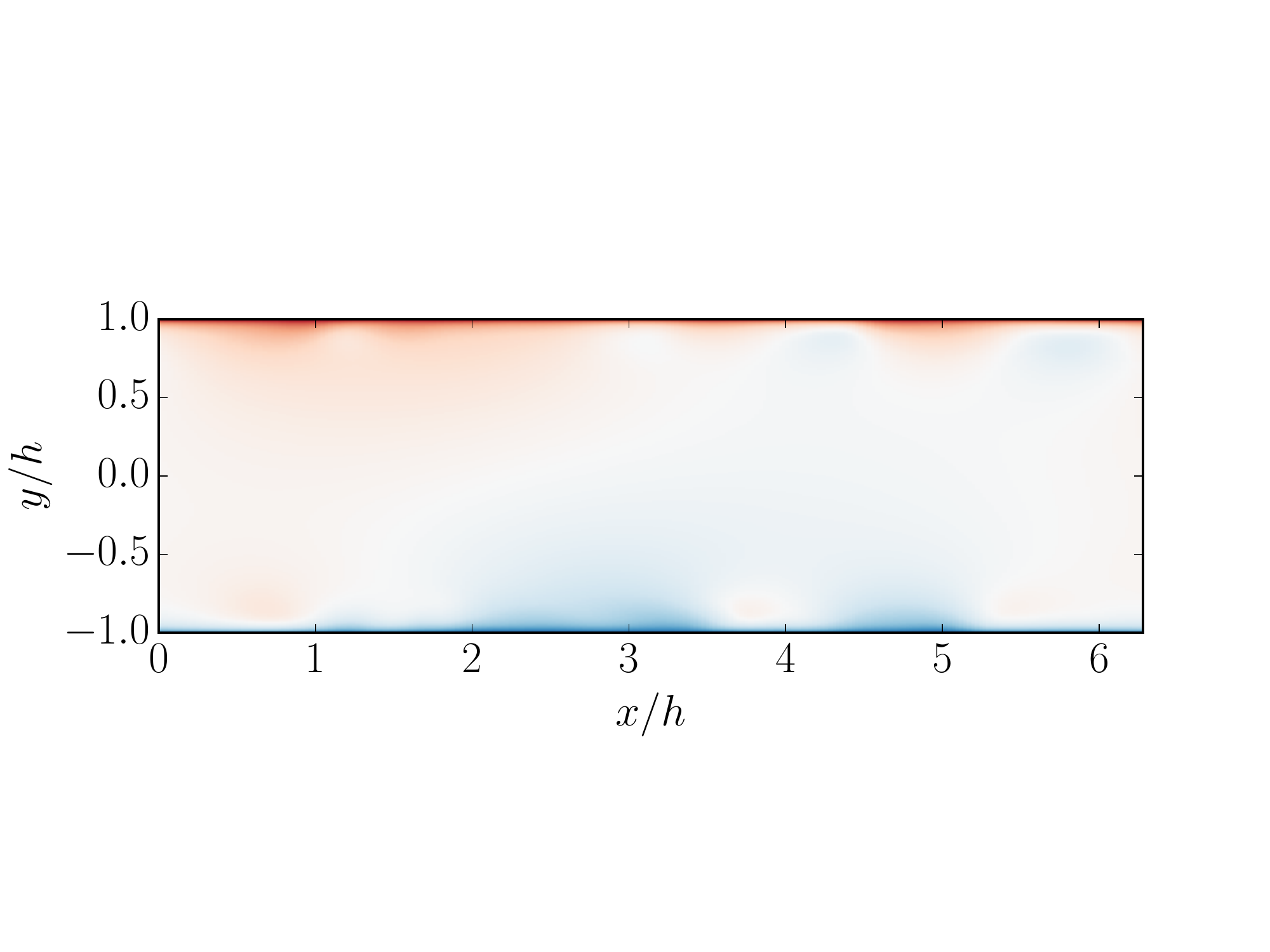}%
  \includegraphics[width=0.45\textwidth,trim={0 3cm 0 3cm},clip]{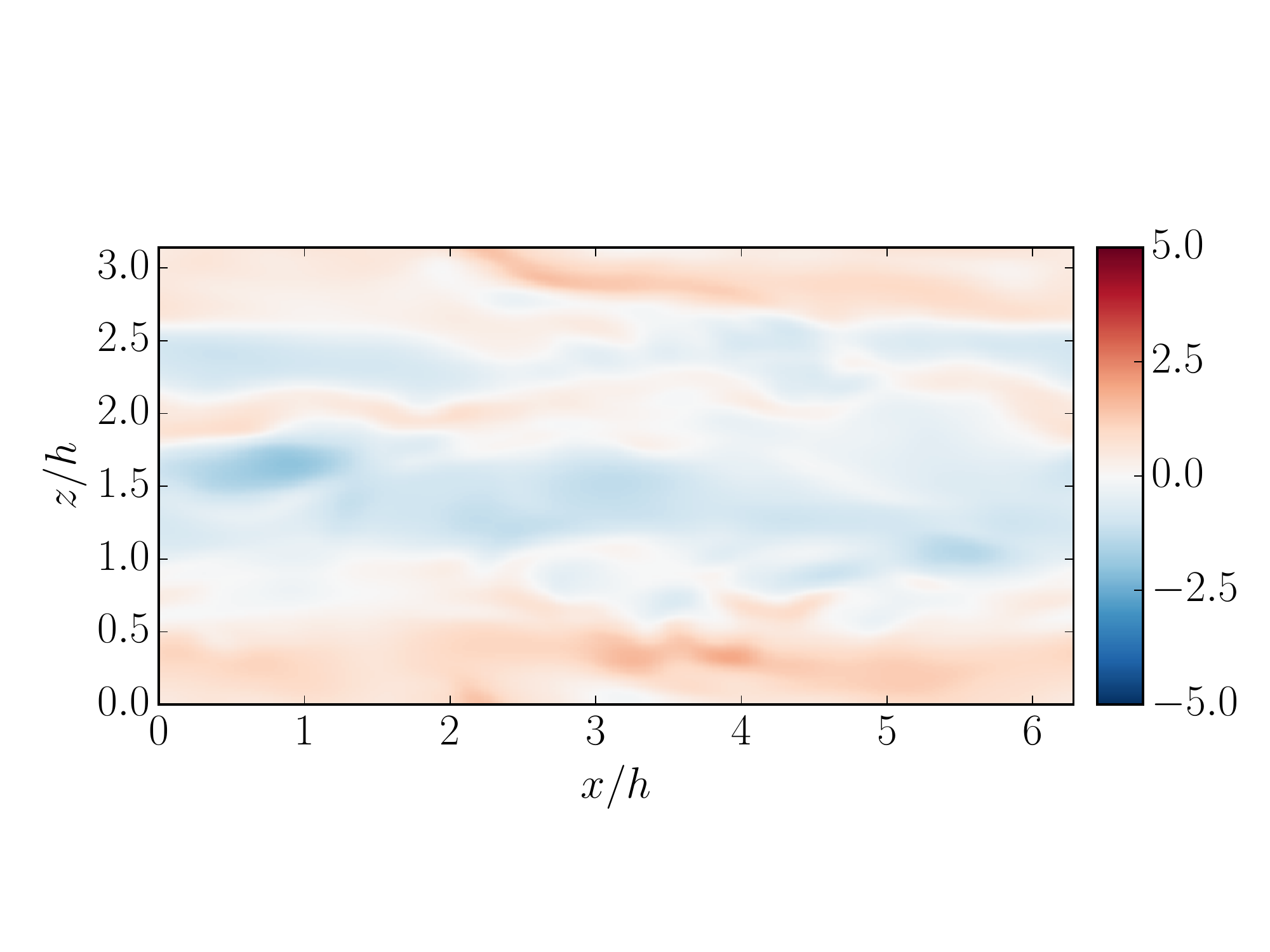}%
\caption{Visualization of the instantaneous electric potential for a streamwise cut (left) and a wall-normal cut (right) inside the buffer layer for the L33 case and $\hat{V}=10$. The color codes for $(\Phi_0+\phi)/V_T$ and the scale is the same for both plots. Comparing the streamwise and wall-normal cuts reveal that the fluctuations are highly anisotropic in character. }
\label{fig:instelp}%
\end{figure}

Figure \ref{fig:rmsphi} shows the RMS fluctuations of $\phi$ for both the L100 and L33 cases. Again, larger fluctuations are seen in the L33 case due to the increased interaction between the electrical double layer and buffer layer. A lingering question is why the scaling estimate Equation (\ref{eq:one}), which predicts that the charge fluctuations should significantly couple back to the flow at the voltages that we have simulated, appears to break down. In Figure \ref{fig:instelp}, we show the instantaneous electric potential $\Phi_0+\phi$. We can see that length scales of the fluctuations are anisotropic. The length scale of fluctuations in the wall-normal direction is $\mathcal{O}(\lambda_d)$, much smaller than the length scale in the horizontal directions (the ones we want to generate to affect the flow). In other words, the gradients generated in the vertical direction are larger than the horizontal direction. Thus, the restoring force for concentration fluctuations in the horizontal direction is high, and fluctuations in the bulk are suppressed. As a consequence this breaks the implicit assumption that the lengthscale of fluctuations in all directions is $\mathcal{O}(\lambda_d)$ when deriving Equation (\ref{eq:one}). Therefore, a significant perturbation to the flow is seen only at voltages higher than that predicted by Equation (\ref{eq:one}).

The results in Figure \ref{fig:results1} suggest that the goal of controlling friction is within reach if we can tune the ion concentration fluctuations to hit the sweet-spot in the buffer layer, and increase the magnitude of the ion concentration and electric potential fluctuations. The magnitude of the concentration fluctuations is small for the panels shown here where $\hat{V}=10$. Increasing the applied voltage $\hat{V}$ beyond $\hat{V}=40$ will result in larger fluctuations in the concentration field. However, for large $\hat{V}$, the large values of the concentration fields near the electrodes are both unphysical and cause numerical instabilities. Steric effects beyond the point ion approximation in the Poisson-Nernst-Planck formalism must be accounted for.

\section{Discussion and Conclusion} 

This manuscript presents a novel code to simulate three dimensional turbulent flows in electrolytes by fully solving the Navier-Stokes equations coupled with the Poisson-Nernst-Planck equations. We have numerically simulated a dilute electrolyte between two electrodes which are sheared, driving a turbulent flow. No statistically significant change in the drag is observed when an electric field is applied. We have not found evidence that the ion perturbations make it into the flow, increasing or decreasing the turbulence level. The largest (but still within statistical error) effect on friction is seen when the region of maximal concentration fluctuations coincides with the location of peak turbulence production of the flow. 

Our results show that the effects of the electrical double layer is insufficient to substantially modify friction in the high Reynolds number regime for the parameters studied. Therefore, the statement posed in the title is answered in the negative. However, importantly our conclusion is only true within the Poisson-Nernst-Planck formalism. As such, possibilities are abound for reversible drag control in regimes where the simple Poisson-Nernst-Planck formalism breaks down.

First, considering higher voltages requires the development of more efficient numerical schemes to solve equations of ion transport which accounts for steric effects close to the walls \cite{kilic2007steric,storey2012effects,kondrat2015dynamics}. The steric effect sets a maximum for the concentration field and thickens the electrical double layer so that it could penetrate deep into the turbulence generating mechanism. 

Moreover, while the Poisson-Nernst-Planck model can capture the existence of a double-layer, it does not capture physics such as the variation in the local viscosity of the fluid as a function of local ion concentration \cite{bazant2009nonlinear}. An ion concentration-dependent viscosity could cause the emergence of a quiescent layer of higher ion concentration, thus higher viscosity, near the wall. Intuitively, this quiescent layer causes two competing effects: on the one hand, the quiescent layer effectively decreases the gap width, which would naturally increase the friction. On the other hand, the presence of this layer could play a role in disrupting the turbulence generation cycle as one would expect that the turbulence is weaker in regions of higher viscosity, thus potentially decreasing friction. This non-trivial confluence of factors provides an avenue for further exploration. 

From a numerical point of view, we have artificially raised the $Sc$ number of the flow to 3 in order to make the simulations feasible. While we argue that this should not play a significant effect, the role of the $Sc$ number requires further numerical verifications. Future studies of both the computational requirements to run high $Sc$ electrolytes and the effect on the physics are required.

Finally, it remains to be seen whether a time dependent or spatially varying potential difference could affect turbulence drag. A related point is the fact that more complex flow fields can be realised in electrolyte systems other then the plane Couette flow. For example, electrovective instabilities can occur near ion-selective surfaces \cite{druzgalski2013direct}. We believe that investigating the effect of those electrokinetic instabilities on shear flows in an exciting direction.

\noindent \emph{Acknowledgments:} AAL acknowledges support from the George F. Carrier Fellowship at Harvard University. We thank M. P. Brenner, R. Verzicco and Y. T. Yang for valuable discussions. We acknowledge computing time from the Dutch Supercomputer Cartesius through an NWO grant.

\appendix

\section{Derivation of equation \ref{eq:one}}

Velocity fluctuations inside the buffer layer will have a characteristic velocity $u_\tau$, and a characteristic length scale $\delta_\nu$. Therefore, we can estimate the inertial forces felt by this fluctuation as:

\begin{equation}
 \rho_w \textbf{u}\cdot\nabla\textbf{u} \sim \rho_w u_\tau^2 / \delta_\nu.
 \label{eq:est1}
\end{equation}

\noindent The inertial force fluctuation generates fluctuations in charge density: we can estimate the potential fluctuation as $V$, and the charge fluctuation $\rho_e$ as:

\begin{equation}
 \rho_e \sim \epsilon V / \lambda_d^2.
\end{equation}

\noindent Substituting this into the equation for the electrical force, we arrive at the estimate:

\begin{equation}
 \rho_e \nabla \phi \sim \epsilon V^2 / \lambda_d^3.
 \label{eq:est2}
\end{equation}

\noindent For the fluctuations in the electric field to affect the flow, they must be of the same order of magnitude or larger than the inertial forces:

\begin{equation}
 \rho_e \nabla \phi \gtrsim \rho_w \textbf{u} \cdot \nabla \textbf{u}
\end{equation} 

\noindent Substituting (\ref{eq:est1}) and (\ref{eq:est2}) for these two forces, we obtain:

\begin{equation}
 \epsilon V^2 / \lambda_d^3 \gtrsim \rho_w u_\tau^2 / \delta_\nu.
\end{equation}

\noindent Assuming $\lambda_d \sim \delta_\nu$, and with some algebraic manipulation, we arrive at the condition:

\begin{equation}
 \beta Sc^{-1} \hat{V}^2 \gtrsim \mathcal{O}(1)
\end{equation}

\bibliography{ref_efield} 
\end{document}